\documentclass[12pt]{report}
\usepackage{epsf,amsmath}
\usepackage{indentfirst}
\usepackage{amssymb}
\usepackage{array,tabularx}
\def\@biblabel#1{#1.}
\makeatother

\begin{document}

\title{\large \bf The Mass of the Black Hole in the X-ray Binary M33 X-7
and the Evolutionary Status of M33 X-7 and IC 10 X-1}

\author{\large\bf M.K. Abubekerov, E.A. Antokhina, A.I. Bogomazov,\\
\large\bf A.M. Cherepashchuk\\
\normalsize\it Sternberg Astronomical Institute, Moscow State University \\
Astronomy Reports, vol. 53, no. 3, pp. 232-242 (2009)}

\date{\begin{minipage}{15.5cm} \small
Abstract: We have analyzed the observed radial-velocity curve for
the X-ray binary M33 X-7 in a Roche model. We have analyzed the
dependence between the component masses and the degree of filling
of the optical star's Roche lobe to obtain the ratio of the masses
of the optical star and compact object. For the most probable mass
of the optical star, $m_v=70M_{\odot}$, the mass of the compact
object is $m_x=15.55\pm3.20 M_{\odot}$. It has been shown that
black holes with masses of $m_x=15M_{\odot}$ and even higher can
form in binaries. We present characteristic evolutionary tracks
for binary systems passing through an evolutionary stage with
properties similar to M33 X-7 - type objects. According to
population-synthesis analyses, such binaries should be present in
galaxies with masses of at least $10^{11}M_{\odot}$. The present
number of such systems in M33 should be of the order of unity. We
have also studied the evolutionary status of the X-ray binary IC
10 X-1 with a Wolf-Rayet component, which may contain a massive
black hole. The final stages of the evolution of the M33 X-7 and
IC 10 X-1 systems should be accompanied by the radiation of
gravitational waves.
\end{minipage}
} \maketitle \rm

\section*{\normalsize INTRODUCTION}
\normalsize

The X-ray source M33 X-7 (which we will call X-7) is situated in
the galaxy M33 X-7 \cite{Long1981}. The periodic variability of
the source was first detected by Peres et
al.\cite{Peres1989a,Peres1989b}, whose gave the first estimate for
the variability period $P_{orb}=1^d.7857$ and suggested that the
X-ray source was in a stellar binary system.

Based on X-ray data obtained by the Einstein Observatory, ROSAT,
and ASCA, Larson and Schulman \cite{Larson1997} refined the period
of the X-ray variability in the binary, $P_{orb}=3^d.4531$. This
value was confirmed by Dubus et al. \cite{Dubus1999}:
$P_{orb}=3^d.4535\pm0.0005$.

In 2004, the X-ray source X-7 was identified with an optical star
of magnitude 18.89m \cite{Pietsch2004}. Photometric observations
in B and V \cite{Pietsch2004} confirmed that X-7 was in a binary,
and showed that the spectral type of the optical component of X-7
is B0I-O7I and its mass is $m_v=25-35M_{\odot}$
\cite{Pietsch2004}. The refined period of the binary (based on
XMM-Newton and Chandra X-ray data) $P_{orb}=3^d.45376\pm0.00021$
was obtained \cite{Pietsch2004}

Based on the Einstein, ROSAT, and XMM Newton data, Pietsch et al.
\cite{Pietsch2006} derived the new period
$P_{orb}=3^d.453014\pm0.000020$. HST observations of the OB
association HS 13, which contains X-7, yielded the refined
spectral type for the optical component O6 III, and a minimum mass
of $m_v=20M_{\odot}$ \cite{Pietsch2006}.

Based on the spectral type of the optical component, the rate of
variation of the binary period, the absence of an X-ray pulsar,
and the analysis of the X-ray spectrum and optical light curves of
\cite{Pietsch2004}, Pietsch et al.\cite{Pietsch2006} suggested
that X-7 is a black hole (BH).

A detailed analysis of the photometric light curves of X-7 is
presented in \cite{Shporer2007}. The photometric period of the
binary was $P_{orb}=3^d.4530\pm0.0014$. The radius of the optical
component is $R=15-20M_{\odot}$, and its effective temperature
$T_{eff}=33000-47000 K$ \cite{Shporer2007}.

The first dynamical estimate of the mass of the compact object in
X-7 was derived from the observed radial-velocity curve in
\cite{Orosz2007}: $m_x=15.65\pm1.45M_{\odot}$. The average mass
for most black-hole candidates in close binary systems is near
$m_x\sim 8M_{\odot}$. The mass of the compact object in X-7
significantly exceeds this estimate. Note that the mass of the
compact object was estimated for a point-mass model in
\cite{Orosz2007}. In addition, the errors of the observed
radial-velocity curve are large (about 20-30\% of the
half-amplitude), which certainly affects the accuracy of the
estimated mass of the compact object. Having in mind the
importance of a reliable mass for the compact object in X-7 for
our understanding of the evolution of binary stars, we have
analyzed the observational data of \cite{Orosz2007} in a Roche
model \cite{Antokhina1994,Antokhina1996}. We have also analyzed
the evolutionary status of the X-7 system and the evolution of the
X-ray binary IC 10 X-1, which contains a Wolf-Rayet (WR) star and
a possible black hole with a mass of $m_x\sim 23M_{\odot}$.

\section*{\normalsize OBSERVATIONAL DATA}

\begin{figure*}
\vspace{0cm} \epsfxsize=0.99\textwidth \epsfbox{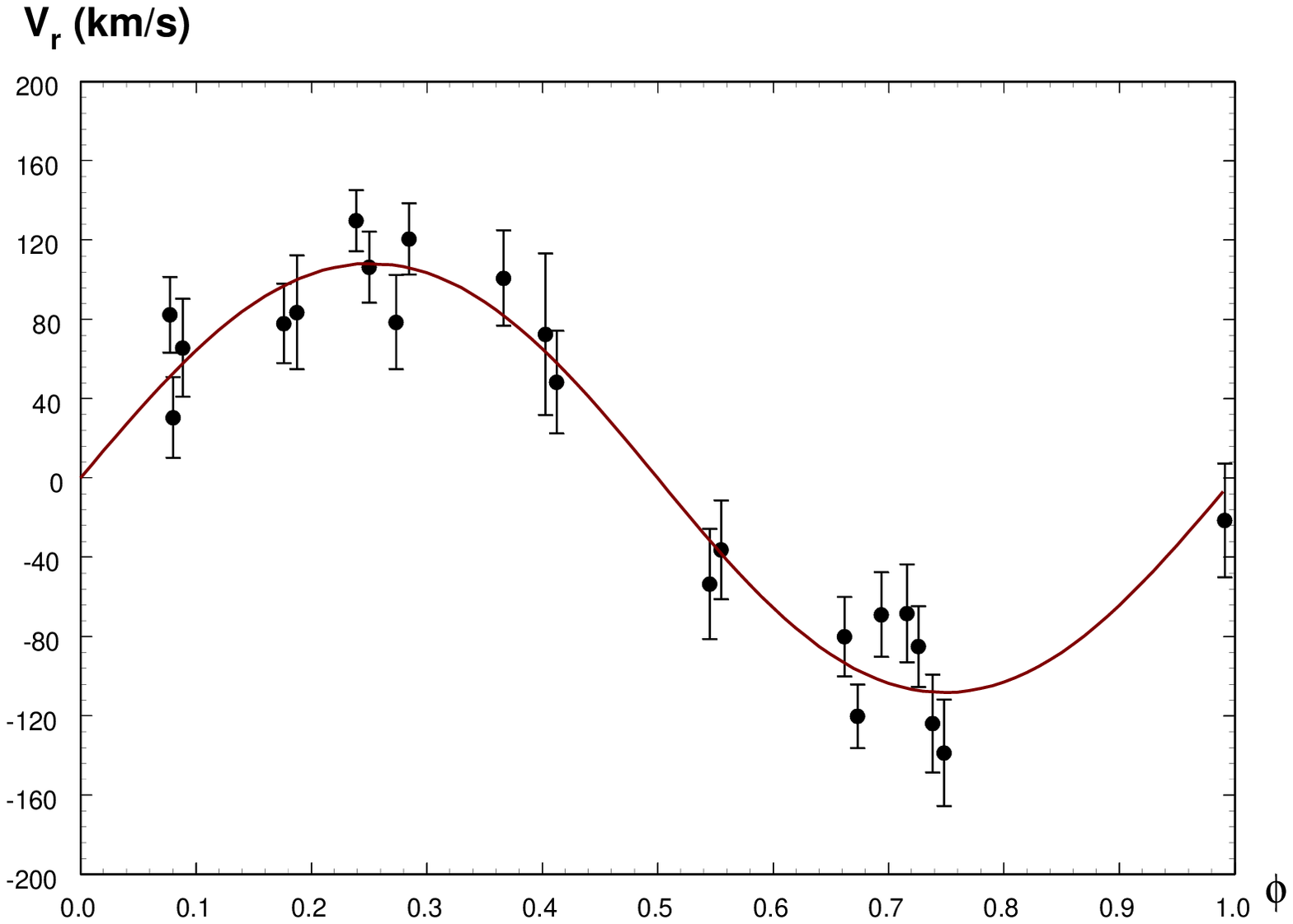}
\caption{Observed and theoretical radial-velocity curve for the
optical component in the X-ray binary M33 X-7. The points show the
radial velocities of the optical star taken from \cite{Orosz2007}.
The solid curve shows the theoretical radial-velocity curve in the
Roche model calculated for a mass of the compact object
$m_x=15.55M_{\odot}$, a mass of the optical star
$m_v=70M_{\odot}$, and the orbital inclination $i=74.6^{\circ}$.
(the remaining parameters of the binary are presented in Table
\ref{param Roche}).} \label{Vobs}
\end{figure*}

We used the spectral data of \cite{Orosz2007} obtained from August
18 to November 16, 2006 on the 8.2-m Gemini North Telescope at
wavelengths $\lambda=4000-5000$ \AA. The radial-velocity curve was
derived from 22 spectra using the cross-correlation method
together with a synthetic spectrum for $\lambda=4150-4300$ \AA\
and $\lambda=4521-4578$ \AA. These wavelength intervals contain
the HeII 4200 \AA\, and HeII 4541 \AA\, lines, unblended by
nebular lines. Orosz et al. \cite{Orosz2007} used the time for the
middle of the X-ray eclipse of X-7 as the zero phase, $T_0 (HJD)=
2453967.157\pm 0.048$ (this is fully consistent with the zero
phase obtained previously in \cite{Pietsch2006}, and differs from
it by $95.001\pm0.014$ orbital cycles). The binary orbital period
$P_{orb}=3^{d}.453014$ \cite{Orosz2007} was obtained from the
X-ray light curve. Figure \ref{Vobs} presents the observed
radial-velocity curve.

Since the data of \cite{Orosz2007} are not presented in numerical
form, we digitized their observed radial-velocity curve with an
accuracy for the radial velocity of the optical component $\sim
0.6$ km/s, which is substantially smaller than the mean error of
the observed radial velocities, $\sim 20-25$ km/s. Therefore, the
accuracy of our digitization of the observed radial-velocity curve
from \cite{Orosz2007} is fully suitable for our further analysis.

\section*{\normalsize ANALYSIS FOR THE OBSERVED
RADIAL-VELOCITY CURVE}

The distance between the components in the X-7 binary system is
comparable to the radius of the optical star. The degree of
Roche-lobe filling by the optical component in X-7 is
$\mu=0.777\pm0.017$ \cite{Orosz2007}. The point-mass model used to
interpret the radial-velocity curve in \cite{Orosz2007} does not
allow taking into account the proximity of the components.

In this connection, we analyzed the radial-velocity curve for the
optical component of X-7 in a Roche model, which enables us in a
first approximation to take into account the tidal deformation of
the star and the non-uniform temperature distribution over its
surface, due to the effects of gravitation darkening and heating
of the star's surface by X-rays from the relativistic object. The
effect of X-ray heating in this system is insignificant. The
bolometric luminosity of the optical component for the adopted
effective surface temperature $T_{eff}=35000$K is
$L_v=1.3\cdot10^{39}$ erg/s. The average luminosity of the X-ray
source out of eclipse is $L_x=5\cdot10^{37}$ \cite{Pietsch2006}.
Thus, $k_x=L_x/L_v\simeq0.04$ (in spite of the small amount of
X-ray heating of the surface of the optical star, we including
this heating when modeling the radial-velocity curve). Therefore,
the radial-velocity curve for the optical component is mainly
perturbed by tidal-rotational deformation of the star and
gravitational darkening of its surface. The procedure used to
model the radial-velocity curve is described in detail by
Antokhina and Cherepashchuk \cite{Antokhina1994} and Antokhina
\cite{Antokhina1996}. Here, we briefly review the main elements of
the method.

The binary system consists of the optical star, described by the
Roche model, and a point-like X-ray source. The tidally deformed
surface of the star was subdivided into $\sim2600$ surface
elements, for each of which the emergent radiation was calculated,
taking into account gravitation darkening, the heating of the
star's surface by radiation from its companion (the reflection
effect), and limb darkening. The effect of heating of the star's
atmosphere by X-ray radiation from the companion was included in
the procedure \cite{Antokhina1994,Antokhina1996} by adding the
emergent and incident fluxes (without taking into account
radiation transfer in the star's atmosphere). Since the effect of
X-ray heating is very small in X-7, this approximation is quite
acceptable.

A model radial-velocity curve was calculated based on the $\text
H_{\gamma}$ absorption line profile. The pro- file and its
equivalent width for each visible surface element with local
temperature $T_{loc}$ and local gravitational acceleration
$g_{loc}$ were calculated by interpolating values in Kurucz's
tables for Balmer lines \cite{Kurucz1979}. The integrated line
profile at a given orbital phase was calculated by summing over
the visible surface of the star, taking into account the Doppler
effect for the local profiles, previously normalized to the
continuum for each surface element (see
\cite{Antokhina1994,Antokhina1996} for more detail). The
calculated integrated absorption profile was used to determine the
radial-velocity of the star. The radial velocity at a given
orbital phase was calculated from the average wavelength at
residual intensity levels of 1/3, 1/2, and 2/3 of the depth of the
integrated absorption profile. The limb darkening of the stellar
disk was calculated using the square-root law
$I(cos\theta)=I_0(1-x(1-\cos\theta)+y(1-\sqrt{\cos\theta}))$,
where $\theta$ is the angle between the direction towards the
observer and the normal to the surface and $x$ and $y$ are the
limb-darkening coefficients \cite{DiazCordoves1992,VanHamme1995}.

\begin{table}[h!]
\caption{Parameters used in modeling the radial-velocity curves
for the optical component in the Roche model.}\label{param Roche}
\vspace{3.0mm} \centering
\def\arraystretch{1.0}
\newcolumntype{C}{>{\centering\arraybackslash}m}
\begin{tabular}{|l|l|p{110mm}|}
\hline
$P$(ñóò.)         & 3.453014         & Orbital period \\
$m_v$($M_{\odot}$)& $var^{*}$        & Mass of the optical star \\
$e$               & 0.0              & Eccentricity \\
$i(^\circ)$       & $74.6$           & Orbital inclination \\
$\mu$             & 0.78-1.0         & Roche-lobe filling factor for the optical component \\
$f$               & 1.0              & Asynchronicity factor for the rotation of the optical component\\
$T_{\text{eff}}$(K)& 35000           & Average effective temperature of the optical component\\
$\beta$           & 0.25             & Gravitational darkening coefficient \\
$k_x$             & $0.04$           & Ratio of the X-ray
                                        luminosity of the relativistic component to the bolometric
                                        luminosity of the optical component
                                        $L_x/L_{v}$ \\
$A  $             & 1.0              & Coefficient for reprocessing of the X-ray radiation \\
$x$; $y$      & $-0.186$; $-0.683$   & Limb darkening coefficient \\

\hline \multicolumn{3}{c}{}\\
\multicolumn{3}{c}{\begin{minipage}{15cm}\small $^{*}$ - The mass
of the optical star was varied in the model calculations; $\mu$
was taken to be 0.78, 0.85, 0.90, 0.95, 1.0.
\end{minipage}}\\
\end{tabular}
\end{table}

Since the mass of the optical star is not known exactly, we
treated the masses of both components of the binary as parameters
of the model to be fit. We carried out an exhaustive search over
the parameters and over multiple solutions for the fit. An
exhaustive search for the mass of the compact object $m_x$ was
made for each mass of the optical component $m_v$ from the
discrete set of values $20M_{\odot}$, $30M_{\odot}$,
$40M_{\odot}$, $50M_{\odot}$, $60M_{\odot}$, $70M_{\odot}$,
$80M_{\odot}$ and for the orbital inclination $i=74^{\circ}.6$.
The fit of the model to the observations was evaluated using the
$\chi^2$ criterion. We selected the significance level
$\alpha=5\%$ (see \cite{Cherepashchuk1993} for more detail).

\begin{figure*}
\vspace{0cm} \epsfxsize=0.99\textwidth
\epsfbox{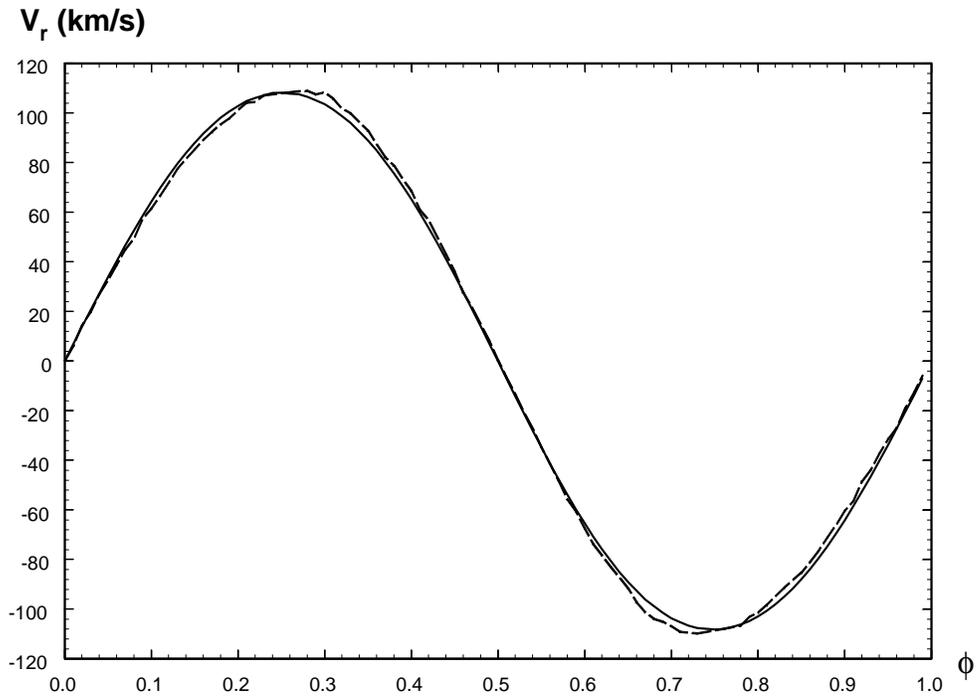} \caption{Theoretical
radial-velocity curves for the optical star, obtained in the Roche
model with the mass of the compact object $m_x=15.55M_{\odot}$,
the mass of the optical star $m_v=70M_{\odot}$, and the orbital
inclination $i=74.6^{\circ}$ (the other parameters of the binary
are presented in Table \ref{param Roche}). The solid and dashed
curves represent the radial-velocity curves calculated from the
$H_{\gamma}$ and HeII 4200\AA\, absorption lines, respectively.}
\label{Vteor_H_HeII}
\end{figure*}

In \cite{Orosz2007}, the observed radial velocities were obtained
from fragments of spectra containing the HeII 4200\AA\ and HeII
4541\AA. Unfortunately, fitting the radial-velocity curve based on
the HeII model line is very time-consuming. Therefore, as we
indicated above, we constructed the theoretical radial-velocity
curve using the model $\text H_{\gamma}$ absorption line. Test
calculations (Fig.\ref{Vteor_H_HeII}) showed no qualitative
difference between the model radial-velocity curves obtained from
the HeII 4200\AA\, and $\text H_{\gamma}$ lines; the average
quantitative difference is 3-4 km/s, which is substantially
smaller than the errors in the observed radial velocities, $\Delta
V_r\simeq 20-30$ km/s. The calculated radial-velocity curve
derived for the HeII 4200\AA\, line was based on the model
atmosphere for the optical star \cite{Antokhina2005}. Figures 3.5
present model profiles of the HeII 4200 \AA\, absorption line at
the two orbital phases $\phi=0.0$, $\phi=0.25$ and of the $\text
H_{\gamma}$ absorption line profile for Roche-lobe filling factors
of $\mu=0.78$, $\mu=1.0$. The result of our analysis of the
observed radial-velocity curve of \cite{Orosz2007} in the Roche
model is presented in numerical form in Table \ref{mvmxtab} and
graphically in Fig.\ref{mx_f_mv}.

Figure \ref{Xi2} presents the discrepancy $\Delta$ between the
observed and theoretical radial-velocity curves, whose minimum
corresponds to the best-fit relationship between $m_v$ and $m_x$.
It is obvious that the Roche model for X-7 cannot be rejected at
the $\alpha=0.05$ significance level. The model can be adopted,
and the optimum values for the parameters and their confidence
intervals (errors) at the $\gamma=1-\alpha=0.95$ confidence level
can be estimated.

\small
\begin{table}[h!]
\caption{Dependence of component masses in X-7 for
$i=74^{\circ}.6$. (the errors correspond to the 95\% confidence
interval for $\chi^2_M$ statistics, where $M$ is the number of
points in the radial-velocity curve).}\label{mvmxtab}
\vspace{3.0mm}\centering
\def\arraystretch{1.4}
\newcolumntype{C}{>{\centering\arraybackslash}m}
\begin{tabular}{|C{15mm}|C{15mm}|C{15mm}|C{15mm}|C{15mm}|C{15mm}|C{15mm}|C{15mm}|}
\hline
 $m_v(M_{\odot})$  &  20            & 30            & 40              & 50             & 60            & 70             & 80              \\
\hline
 $m_x(M_{\odot})$  &  $7.30\pm1.60$ & $9.30\pm2.00$ & $11.0\pm2.40$   & $12.65\pm2.70$ & $14.15\pm2.90$ & $15.55\pm3.20$ & $16.95\pm3.40$  \\
\hline
\end{tabular}
\end{table}
\normalsize

Our analysis using our more complex and physically more adequate
model confirms the quantitative conclusions of \cite{Orosz2007}.
Due to the small degree of Roche-lobe filling by the optical star,
the black-hole masses obtained in our study and in
\cite{Orosz2007} are quite similar: $15.55\pm3.20M_{\odot}$ and
$15.65\pm1.45M_{\odot}$, respectively (for the most probable mass
of the optical star $m_v=70M_{\odot}$).

Let us consider the dependence of the amplitude of the theoretical
radial-velocity curve on the degree of Roche-lobe filling by the
optical star. Figures \ref{HeIIph025mu078mu1} and
\ref{Hgph025mu078mu1} present the model HeII 4200 \AA\ and $\text
H_{\gamma}$ absorption line profiles obtained for degrees of
Roche-lobe filling by the optical star $\mu=0.78$ and 1.0. The
"center of gravity"\, for the absorption line obtained for
$\mu=0.78$ is more shifted towards red wavelengths, compared to
the position for complete filling of the Roche lobe; i.e., the
half-amplitude of the radial-velocity curve of the optical star
$K_v$ decreases with increasing degree of Roche-lobe filling
$\mu$.

The following calculations present the quantitative difference
between the half-amplitude of the radial-velocity curves for the
point-mass and Roche models for various degrees of Roche-lobe
filling $\mu$. For these calculations, we took the masses of the
compact and optical components to be $m_x=15.55M_{\odot}$ and
$m_v=70M_{\odot}$. The half-amplitude of the radial-velocity curve
for these parameters of the binary was $V_c=108.65$ km/s. The
difference of the half-amplitude of the radial-velocity curve
presented in Table \ref{mu_Vr} is essentially independent of the
orbital inclination of the binary (at least in the interval
$i=60^{\circ}-90^{\circ}$). It is obvious that the half-amplitude
of the theoretical radial-velocity curve decreases with increasing
$\mu$, and is 6.11 km/s smaller than the value for the point-mass
model for $\mu = 1.0$. The mass of the relativistic object $m_x$
in the Roche model increases with increasing $\mu$.

\begin{table}[h!]
\caption{Difference between the half-amplitude of the
radial-velocity curves obtained in the Roche and point-mass
models, $V_{Roche}$ and $V_c$, as a function of the degree of
Roche-lobe filling by the optical star $\mu$.} \label{mu_Vr}
\vspace{3.0mm} \centering
\def\arraystretch{1.0}
\newcolumntype{C}{>{\centering\arraybackslash}m}
\begin{tabular}{|C{35mm}|C{15mm}|C{15mm}|C{15mm}|C{15mm}|C{15mm}|}
\hline
 $\mu$                       &  0.78    & 0.85   & 0.90    & 0.95    & 1.00   \\
\hline
 $V_c-V_{Roche}$ (êì/ñ)      &  0.80    & 1.43   & 2.11    & 3.24    & 6.11   \\
\hline $m_x(M_{\odot})^{*}$  &  15.55   & 15.60  & 15.65   & 15.75   & 15.95  \\
\hline\multicolumn{6}{c}{}\\
\multicolumn{6}{c}{\begin{minipage}{13cm}\small $^{*}$ -
Black-hole mass obtained in the Roche model for the mass of the
optical star $m_v=70M_{\odot}$. The remaining model parameters are
presented in Table \ref{param Roche}.
\end{minipage}}\\
\end{tabular}
\end{table}

We can see from Table \ref{mu_Vr} that, up to $\mu=0.90$, the
difference between the half-amplitudes of the radial-velocity
curves is close to $\sim 1\%$ of the half-amplitudes themselves.
When the Roche-lobe filling is even higher, $\mu=0.95-1.0$, the
half-amplitude of the radial-velocity curve decreases by $\sim
3-6\%$ compared to the half-amplitude for the point-mass model.

The results of fitting the observed radial-velocity curve for X-7
in the Roche model are close to those obtained for the point-mass
model, even when the Roche lobe of the optical star is fully
filled. For example, the difference between the mass of the
compact object obtained in the Roche model for $\mu=0.78$ and in
the point mass model is 0.6\%, while this difference is 2.5\% for
$\mu=1.0$. This closeness of the estimates for the mass of the
compact object obtained in the Roche and point-mass models is not
universal. For example, when a point-mass model is used to
estimate the masses of X-ray pulsars in binaries with OB giants,
the masses of the compact objects are systematically
underestimated by up to $\sim10\%$ (see \cite{Abubekerov2004a,
Abubekerov2004b, Abubekerov2004c}). In the presence of substantial
X-ray heating of the optical star, which appreciably changes the
shape of the absorption profiles, the use of a point-mass model
yields unrealistic results \cite{Abubekerov2006}.

The similarity of the estimates for the mass of the compact object
obtained in the Roche and point mass models for the binary X-7 is
due to the fairly high mass of the compact object and the low
degree of Xray heating, $k_x=0.04$. Thus, the fit of the observed
radial-velocity curve for the optical star in X-7 obtained for our
more adequate model fully confirms the quantitative conclusions
drawn in \cite{Orosz2007}.

In spite of the fact that estimates of the mass of the optical
component vary from $20M_{\odot}$ to $70M_{\odot}$, we have
adopted $m_v=70M_{\odot}$ as the most probable value, since this
is the value obtained from the most complete set of observational
data: a spectral analysis, fitting of the light curve of the
optical star, and the duration of the X-ray eclipse
\cite{Orosz2007}.

Based on the mass estimate $m_v=70M_{\odot}\pm6.9M_{\odot}$
\cite{Orosz2007}, we find that the mass of the compact object is
$m_x=14.5-16.5M_{\odot}$. As was already noted, this black-hole
mass deviates substantially from the average value $m_x\simeq
8M_{\odot}$ for most binaries with black holes. We also emphasize
the Roche model fit to the radial-velocity curve can be considered
acceptable based on the $\chi^2_M$ criterion at the $\gamma=95\%$
confidence level (where $M$ is the number of observed points in
the radial-velocity curve). Further, we consider evolutionary
scenarios that can result in the formation of systems similar to
X-7.

Recently \cite{Silverman2008}, the minimum masses of the
components of the binary IC 10 X-1, which consists of a Wolf-Rayet
star and a compact object (proposed black hole) were estimated.
The mass of the Wolf-Rayet star is $m_v=32.7\pm2.6M_{\odot}$, and
the mass of the black hole obtained from the observed
radial-velocity curve of the optical component is
$m_x=23.1\pm2.1M_{\odot}$. Since no exact correspondence in time
has been found between the eclipses of the components and the
transition of the radial-velocity curve derived from the HeII
4200\AA, emission line through the $\gamma$-velocity, the
conclusion that the black hole in this system has a high mass must
considered tentative. Nevertheless, we investigated the possible
formation of a binary with these component masses.

\section*{\normalsize EVOLUTIONARY SCENARIOS FOR X-7
AND IC 10 X-1}

The binary X-7 is considered from an evolutionary point of view in
\cite{Orosz2007}. It was concluded that current evolutionary
scenarios fail to explain the existence of systems similar to X-7.
The problem encountered by some scenarios is as follows. Since the
size of the initially more massive star exceeds the current
distance between the components, the system should undergo a
common-envelope stage, during which the components approach each
other. However, for the formation of a black hole with amass of
about $15M_{\odot}$ the star should fill its Roche lobe
immediately after the completion of the helium burning in the core
(followed by the common-envelope stage). The radius of the star
should then be smaller than during core burning, and so the Roche
lobe should not be filled.

We analyzed possible evolutionary tracks for X-7 and IC 10 X-1
using the ``Scenario Machine'' code. The principles behind this
code have been described repeatedly, and here we indicate only the
basic parameters used in the calculations. A detailed description
of the ``Scenario Machine'' can be found in
\cite{Lipunov1996,scm2}.

The mass functions for black holes in X-ray binaries were
calculated for various evolutionary scenarios in
\cite{bogomazov2005}. One of the scenarios considered in
\cite{bogomazov2005} predicted the existence of black holes with
masses up to $50 M_{\odot}$. This scenario satisfied the most
important criterion for realistic evolutionary scenarios: the
requirement that there exist Cyg X-1-type systems in the Galaxy.

A set of evolutionary scenarios A, B, C, and W are used in
\cite{bogomazov2005}\footnote{Note that scenarios A, B, C, and W
represent various assumptions concerning the physical parameters
of stars at various stages of their evolution, rather than
specifications for types of mass exchange corresponding to the
classification introduced by Kippenhahn and Weigert
\cite{kippenhahn1967a}.}, which include various mass-loss rates,
dependences of the core mass on the initial mass of the star, and
``mass–radius'' dependences. These scenarios are described in
detail in \cite{scm2,bogomazov2005}. It was shown in
\cite{bogomazov2005} that scenario B must be rejected, since it
cannot explain the existence of a Cyg X-1- type system in the
Galaxy (that is not a statistical deviation). Therefore, we do not
consider this evolutionary scenario here. Note that scenario B is
based on the evolutionary assumptions presented in
\cite{Orosz2007}.

Let us consider the calculated evolutionary tracks for a binary
system for evolutionary scenarios A, C, and W. The latter two
scenarios have a higher massloss rate than scenario A. In a first
rough approximation, they can be taken to correspond to higher
metallicities of the stars. The calculations indicate that X-7 or
IC 10 X-1-type systems cannot be obtained from scenarios C and W.
We have identified a domain of parameters in scenario A for which
the formation of X-7 or IC 10 X-1-type systems appeared to be
possible.

A more detailed description of evolutionary scenario A is given in
\cite{Lipunov1996,scm2,bogomazov2005}. and here we only describe
the parameters used as free parameters of the problem. In our
modeling of possible binary precursors for X-7 and IC 10 X-1, the
fraction and rate of mass loss in the main-sequence and Wolf–Rayet
stages, as well as the fraction of the mass of the pre-supernova
that was enclosed by the event horizon, were varied. The star's
mass-loss rate $\dot M$ is very important for two reasons: it
substantially affects both the semimajor axis of the binary and
the mass of the star. The mass loss rate $\dot M$ in the
main-sequence stage is described by the classical formula

\begin{equation}
\dot M=\frac{\alpha L}{cV_{\infty}}, \label{mlossa1}
\end{equation}

\noindent where $L$is the luminosity of the star, $V_{\infty}$ the
velocity of its wind at infinity, $c$ the speed of light, and
$\alpha$ a free parameter. In scenario A, the change in the mass
$\Delta M$ during a single evolutionary stage does not exceed
$0.1(M-M_{core})$, where $M$ is the mass of the star at the
beginning of the stage and $M_{core}$ is the mass of the stellar
core. We parametrized the mass loss at the Wolf–Rayet stage as
$\Delta M_{WR}=k_{WR}M_{WR}$, where $M_{WR}$ is the maximum mass
of the star at this stage.

The mass of the black hole $M_{BH}$ formed as a result of the
explosion of a pre-supernova with mass $M_{preSN}$ was calculated
from the formula

\begin{equation}
M_{BH}=k_{bh}M_{preSN}, \label{k_BH}
\end{equation}

\noindent where the factor $k_{bh}$ is the fraction of the mass of
the pre-supernova enclosed by the event horizon during the
collapse.

The factors $\alpha$ and $k_{WR}$ took the values 0.3 (the
standard value used for calculations in scenario A) and 0.1. The
weak wind roughly imitates the metallicity of X-7, which is
smaller than the solar value \cite{Orosz2007}; the galaxy IC 10 is
also poor in metals (see, for example, \cite{massey2007a}).
$k_{bh}$ was varied from 0.1 to 1.0.

In the course of the population synthesis, we took the X-7 system
to be a binary that contains a black hole with mass
$m_{BH}=14-17M_{\odot}$ and a main sequence star at the end of its
evolution with mass $m_{v}=65-75M_{\odot}$; the orbital period of
the system is $P_{orb}\lesssim 5^{d}$. The IC 10 X-1 system is
represented in the calculations as a binary containing a black
hole with mass $m_{BH}=23-34M_{\odot}$ and a Wolf–Rayet star with
mass $m_{WR}=17-35M_{\odot}$; the orbital period is
$P_{orb}\lesssim 1.5^{d}$. A population synthesis for $10^6$
binaries was carried out for each set of parameters of the
evolutionary scenario. The results suggest that X-7- type binaries
should exist in galaxies with masses of at least
$10^{11}M_{\odot}$ and a star-formation rate specified by the
Salpeter function. The number of such systems should currently be
of the order of unity (Fig. \ref{num}). Figures
\ref{num}-\ref{evolic10} present the results for the calculations.

Figures \ref{num} and \ref{numic10} show the number of X-7-type
and IC 10 X-1-type systems expected to occur in a spiral galaxy
with a mass of $10^{11}M_{\odot}$ and a star formation rate
specified by the Salpeter function. The numbers in the plots
indicate curves calculated with {\it (1)} $\alpha=0.3$ and
$k_{WR}=0.3$, {\it (2)} $\alpha=0.1$ and $k_{WR}=0.1$, and {\it
(3)} $\alpha=0.1$ and $k_{WR}=0.3$. Figure \ref{num} shows that
there exist values for $k_{bh}$ for which X-7-type systems can be
obtained. Taking into account that the star-formation rate, or at
least the star-formation rate per square area, in M33 exceeds the
average for spiral galaxies of the Local Group (see, for example,
\cite{gardan2007a}), the existence of X-7 can be explained in
scenario A. It is clear from Fig. \ref{numic10} that IC 10
X-1-type systems can obtained for a broad range of values of
$k_{bh}$. Since the galaxy IC 10 has active star formation, we
expect that the rate of formation of massive stars in IC 10 is to
order of magnitude comparable with this rate in the Milky Way (in
spite of the fact that the mass of IC 10 is approximately two
orders of magnitude lower than that of the Milky Way). A burst of
star formation apparently started about 10 million years ago in IC
10 and is still ongoing (see, for example,
\cite{massey2007a,vacca2007a}).

Figure \ref{evol} presents a characteristic evolutionary scenario
resulting in the formation of an X-7-type system. In all the
scenarios in which such a system can be obtained, the qualitative
scheme for the evolution of the binary is very similar. At the
beginning of the evolution, this is a massive close binary, with a
mass of the primary $m_1=80-120M_{\odot}$, a mass of the secondary
$m_2=40-60M_{\odot}$, and an initial semimajor axis
$a\lesssim100R_{\odot}$. The approximate intervals for the initial
parameters of the precursors of X-7-type systems are also shown in
Figs. \ref{m-a} and \ref{m-q}. After the end of core hydrogen
burning, the more massive primary fills its Roche lobe. Matter
starts to flow to the secondary, as a rule, faster than on the
nuclear timescale (stage III in \cite{scm2}); the final part of
the mass transfer can sometimes occur on close to this timescale
(stage IIIe in \cite{scm2}). When the primary has lost its
envelope, it turns into a Wolf– Rayet star, which explodes and
forms a black hole. The weak stellar wind in the main sequence
stage, and also ``correct selection'' of the mass fraction of the
pre-supernova that is enclosed by the event horizon during the
formation of the black hole, make it possible to form a black hole
of the required mass (around $15-20M_{\odot}$) in a sufficiently
close system. Note also that an X-7-type system does not undergo a
common-envelope stage before its formation\footnote{See
\cite{scm2} for more detail. The ``Scenario Machine'' assumes that
type B systems, according to the classification of Kippenhahn and
Weigert, form a common envelope if $q\le q_{cr}=0.3$. Here,
$q=M_{accretor}/M_{donor}$ is the ratio of the masses of the
primary (accretor) and secondary (donor) stars. For the track
presented in Fig. \ref{evol} $q\approx 0.35$ at the beginning of
the stage of Roche-lobe filling of the initially more massive
primary star.}. Qualitatively, the subsequent evolution of an X-7-
type binary is as follows. The secondary (initially less massive)
star fills its Roche lobe, leading to a stage of supercritical
accretion onto the black hole; a common-envelope then forms, and
the binary ends its existence with the merger of its components
and the formation of a Thorne-Zhytkov object. The final result of
the evolution of an X-7-type system is a single massive black
hole. When the black hole moves along a spiral trajectory into the
center of the non-degenerate star (the Thorne-Zhytkov object
stage), the system can be a source of gravitational waves
\cite{Nazin1995}.

Let us now consider the characteristic evolutionary track of a
binary that results in the formation of an IC 10 X-1-type system
(Fig. \ref{evolic10}). At the beginning of the evolution, the mass
of the primary is $m_1=80-120M_{\odot}$, the mass of the secondary
is $m_2=15-60M_{\odot}$, and the initial semimajor axis of the
binary orbit is $a \simeq 170-200 R_{\odot}$. The approximate
intervals of the initial parameters of precursors of IC 10
X-1-type systems are also shown in Figs. \ref{m-a} and \ref{m-q}.
Note that, since the mass of a star can exceed its initial mass
due to mass transfer, the minimum initial mass of the secondary
can be lower than the estimated mass for the Wolf–Rayet star in IC
10 X- 1. After the depletion of hydrogen in its core, the more
massive primary fills its Roche lobe. Matter starts to flow to the
secondary, as a rule, faster than on the nuclear timescale (stage
III in \cite{scm2}); the final part of the mass transfer occurs on
a timescale close to the nuclear timescale (stage IIIe in
\cite{scm2}). When the primary has lost its envelope, it turns
into a Wolf–Rayet star, which explodes and forms a black hole.
Further, the secondary finishes its evolution on the main sequence
and fills its Roche lobe. After a phase of supercritical accretion
onto the black hole during the Roche-lobe filling stage, a
common-envelope stage begins, during which the stars approach each
other very closely and the non-relativistic component loses its
envelope. A binary consisting of a black hole and a Wolf–Rayet
star is formed. In the process of collapse of the Wolf–Rayet star
the gamma-burst may occur in such a binary. The evolution of an IC
10 X-1- type binary ends with the merging of the two black holes
(remnants of the evolution of the system components), driven by
gravitational-wave radiation, resulting in the formation of a
single massive black hole. The merger should be accompanied by a
burst of gravitational-wave radiation.

\section*{\normalsize CONCLUSION}

We have used the radial-velocity curve for the optical star in the
M33 X-7 system \cite{Orosz2007} and parameters of the system
derived from an analysis of the light curve \cite{Orosz2007}
($\mu=0.78$, $i=74^{\circ}.6$) to construct the relationship
between the masses of the black hole $m_x$ and the optical star
$m_v$ (Table \ref{mvmxtab}, Fig. \ref{mx_f_mv}) in a Roche model
for the optical star. We also studied the effect of tidal
deformation and gravitational darkening of the optical star on the
mass of the black hole $m_x$ for various degrees of filling of the
Roche lobe of the optical star $\mu$ (Table \ref{mu_Vr}). For
$\mu=0.78$, the impact of the proximity of the components on the
radial-velocity curve of the optical star is relatively small,
justifying the analysis used by Orosz et al. \cite{Orosz2007} to
derived the black-hole mass $m_x=15.65\pm1.45M_{\odot}$ (for
$m_v=70M_{\odot}$). Our estimated mass for the black hole for
$m_v=70M_{\odot}$ coincides with this value within the errors:
$m_x=15.55\pm3.20M_{\odot}$; the error corresponds to the 95\%
confidence interval for $\chi^2_M$ statistics, where $M$ is the
number of data points in the radial-velocity curve. We stress
that, if the degree of Roche-lobe filling by the optical star
$\mu$ were close to unity, with all other factors being the same,
the black-hole mass would be $m_x=15.95\pm3.2M_{\odot}$, which
exceeds the value for $\mu=0.78$ by $\sim 3\%$. Therefore, it is
important to use a more realistic Roche model that takes into
account the tidal–rotational deformation of the optical star and
its gravitational darkening, rather than a point mass model, when
fitting the radial-velocity curve.

We also considered the evolutionary status of the X-ray binaries
M33 X-7 and IC 10 X-1, and have shown that the existence of such
systems, which probably contain massive stellar black holes, is
quite possible in the galaxies M33 and IC 10 for reasonable
physical assumptions about the initial parameters of these systems
and the mass transfer in them. In the evolution process the IC
10X-1 system should pass through the stage, during which it will
consist of a black hole and the Wolf–Rayet star. At the instant of
collapse of the Wolf–Rayet star the gamma-burst may occur in this
binary. We conclude that the evolution of the IC 10 X-1 system
should end with a merger of two massive black holes, which should
result in a burst of gravitational-wave radiation. In the case of
the X-7 system, a burst of gravitational wave radiation should be
observed during the formation of a Thorne–Zhytkov object
\cite{Nazin1995}. These conclusions are important for
observational programs to be undertaken with modern laser
gravitational-wave telescopes, such as LIGO, VIRGO, etc.

Note that the detection of a black hole in M33 X-7
\cite{Orosz2007} has opened the era of searches for black holes in
other galaxies of the Local Group. This has become possible due to
the commissioning of large, 8-10-m, new-generation telescopes. The
first detections of new stellar-mass black holes in other galaxies
are consistent with the hypothesis that there is a bimodal mass
distribution for relativistic objects
\cite{Cherepashchuk2001}-\cite{Postnov2003}. This distribution and
the presence of a dip in the mass distribution for neutron stars
and black holes in the range $m_x=2-4M{\odot}$
\cite{Cherepashchuk2001} were first deduced from the masses of
about 40 objects ($\sim20$ neutron stars and $\sim20$ black
holes). The detection of new black holes in other galaxies will
make it possible to substantially increase the number of mass
estimates for relativistic objects, and thus to improve the
significance of corresponding statistical conclusions. The
increase in the number of detected black holes in X-ray binaries
will also open fresh opportunities for verifying new theories of
gravitation \cite{Postnov2003,Johannsen2008}.

\section*{ACKNOWLEDGMENTS}

The authors thank V. V. Shimanskii for his assistance in the
calculations and useful discussions, as well as A. V. Tutukov for
valuable comments. The work was supported by the Russian
Foundation for Basic Research (project code 08-02-01220), the
State Program of Support for Leading Scientific Schools of the
Russian Federation (grant NSh- 1685.2008.2), a Presidential Grant
of the Russian Federation to Support Young Russian Candidates of
Science (MK-2059.2007.2), and the Analytical Departmental Targeted
Program ``The Development of Higher Education Science Potential''
(grant RNP- 2.1.1.5940).

\newpage

\renewcommand{\figurename}{Fig.}

\renewcommand{\figurename}{Fig.}

\renewcommand{\figurename}{Fig.}
\begin{figure*}
\vspace{0cm} \epsfxsize=0.99\textwidth
\epsfbox{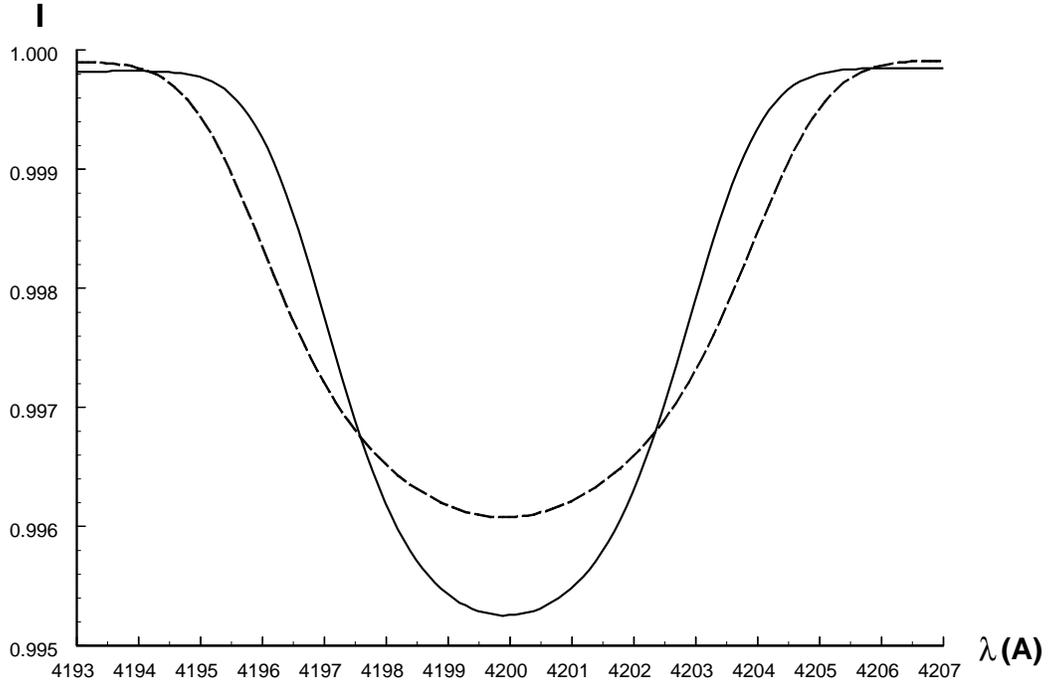} \caption{Model
integrated profiles for the HeII 4200 \AA\, absorption line of the
optical component in X-7 at orbital phase 0.0. The profiles were
obtained in the Roche model with $m_x=15.55M_{\odot}$,
$m_v=70M_{\odot}$, and $i=74^{\circ}.6$. The solid and dashed
curves represent the absorption-line profiles obtained in the
Roche model for degrees of filling of the Roche lobe $\mu=0.78$
and $\mu=1.0$, respectively (the other parameters of the binary
are presented in Table \ref{param Roche}). Both model integrated
profiles were convolved with a spectrograph response function with
FWHM=2\AA.} \label{HeIIph000mu078mu1}
\end{figure*}

\renewcommand{\figurename}{Fig.}
\begin{figure*}
\vspace{0cm} \epsfxsize=0.99\textwidth
\epsfbox{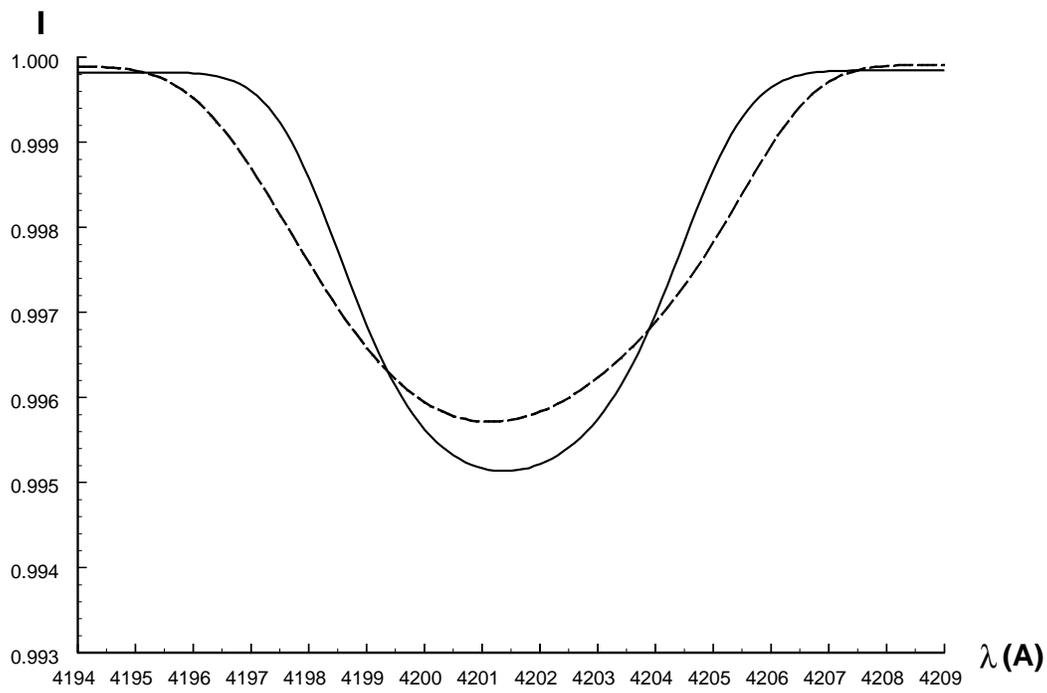} \caption{Same as
Fig.\ref{HeIIph000mu078mu1} for orbital phase 0.25.}
\label{HeIIph025mu078mu1}
\end{figure*}

\renewcommand{\figurename}{Fig.}
\begin{figure*}
\vspace{0cm} \epsfxsize=0.99\textwidth
\epsfbox{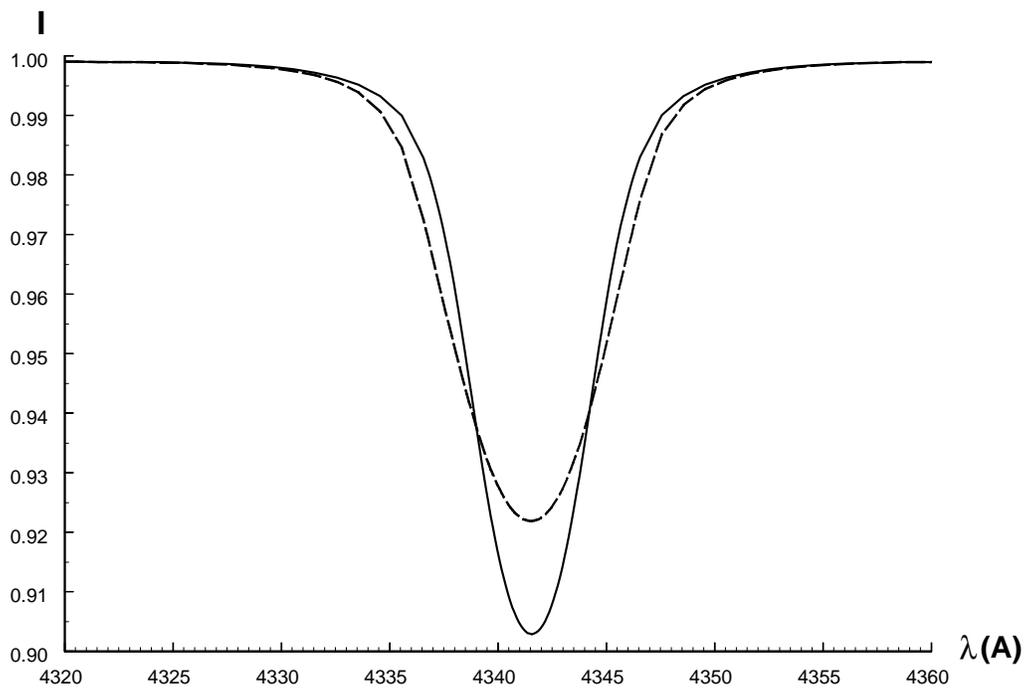} \caption{Same as
Fig.\ref{HeIIph025mu078mu1} for the $\text H_{\gamma}$ absorption
line.} \label{Hgph025mu078mu1}
\end{figure*}

\renewcommand{\figurename}{Fig.}
\begin{figure*}
\vspace{0cm} \epsfxsize=0.99\textwidth
\epsfbox{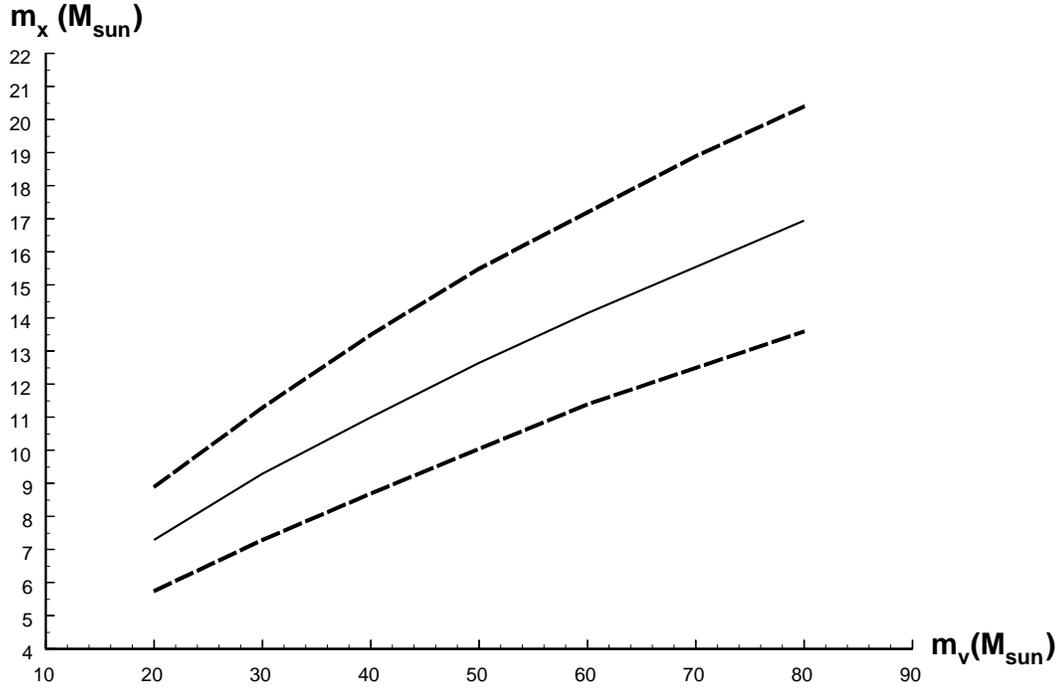} \caption{ Relationship between
the masses of the compact object and optical component of X-7,
obtained from the fit of the observed radial-velocity curve from
\cite{Orosz2007} in the Roche model. The parameters of the binary
are presented in Table 1. The solid curve represents the central
values for the mass of the black hole, while the dashed curves
correspond to the upper and lower boundaries for the errors on the
central values based on the $\chi_M^2$ statistics for the
$\gamma=95\%$ confidence level (M is the number of observed points
on the radial-velocity curve).} \label{mx_f_mv}
\end{figure*}

\renewcommand{\figurename}{Fig.}
\begin{figure*}
\vspace{0cm} \epsfxsize=0.99\textwidth \epsfbox{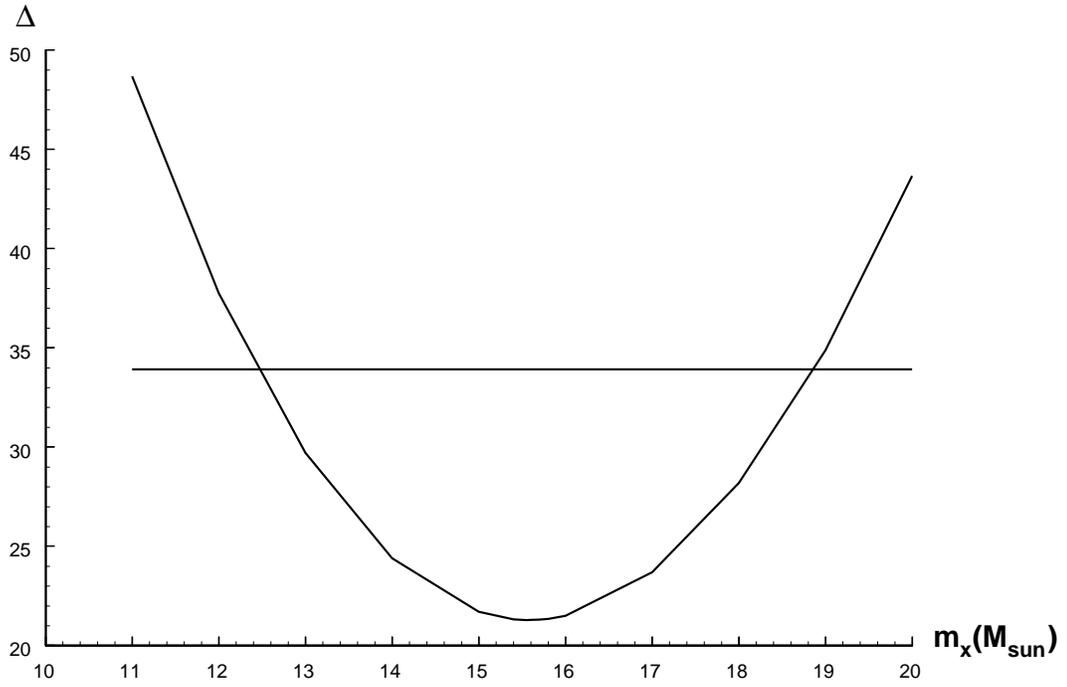}
\caption{Dependence of the discrepancy $\Delta$ between the
observed and theoretical radial-velocity curve for
$m_v=70M_{\odot}$. The horizontal line corresponds to the critical
value for $\chi^2_M$ at the $\alpha=0.05$ significance level. }
\label{Xi2}
\end{figure*}

\begin{figure*}[h!]
\hspace{0cm} \epsfxsize=0.9\textwidth\centering
\epsfbox{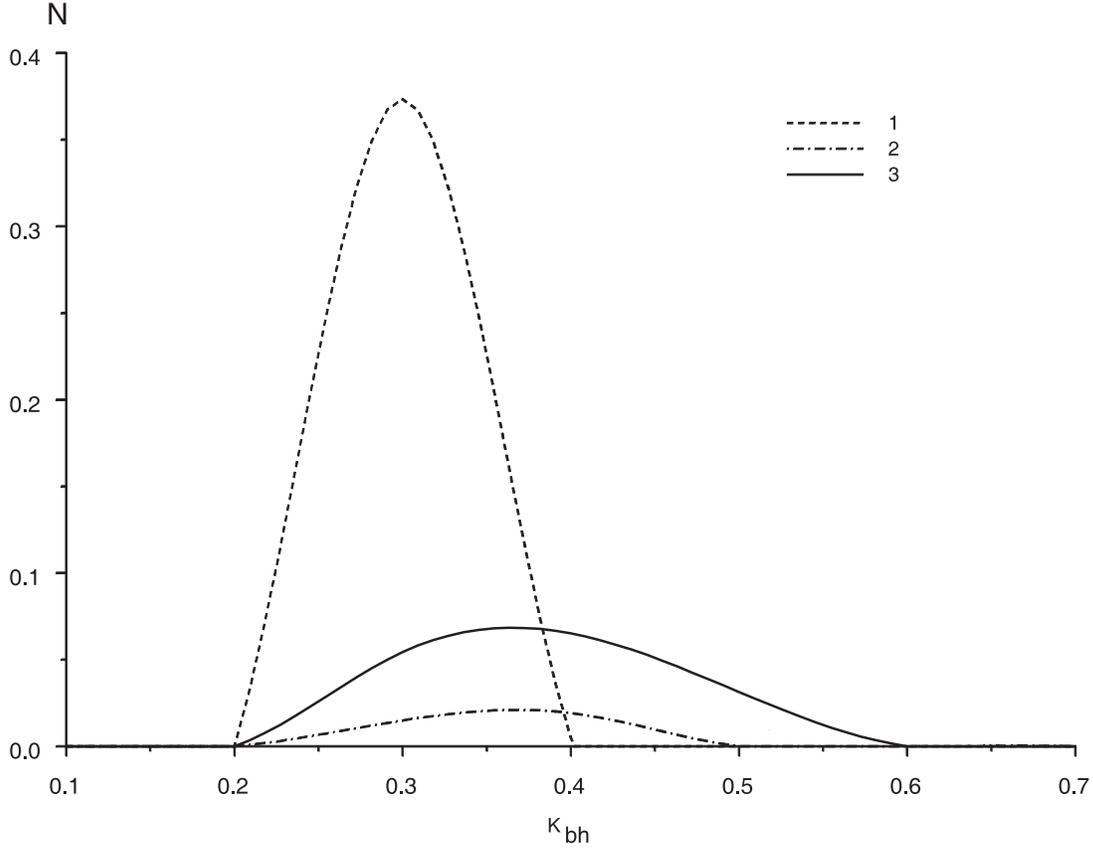} \vspace{0cm}\caption{Number of M33
X-7-type systems for a spiral galaxy with mass $10^{11}M_{\odot}$
and a star-formation rate specified by the Salpeter function. The
numbers indicate curves calculated for various sets of
evolutionary parameters (see the text). The vertical axis plots
the expected absolute number of M33X-7- type systems in the
Galaxy. The horizontal axis plots the fraction of the mass of the
pre-supernova that is enclosed by the event horizon at the instant
when the black hole is formed. In the case of curve ``1''
($\alpha=0.3$, $k_{WR}=0.3$, $k_{BH}=0.3$), the maximum expected
number of M33 X-7-type systems is 0.37, which is of the order of
magnitude of unity.} \label{num}
\end{figure*}

\begin{figure*}[h!]
\hspace{0cm} \epsfxsize=0.5\textwidth\centering
\epsfbox{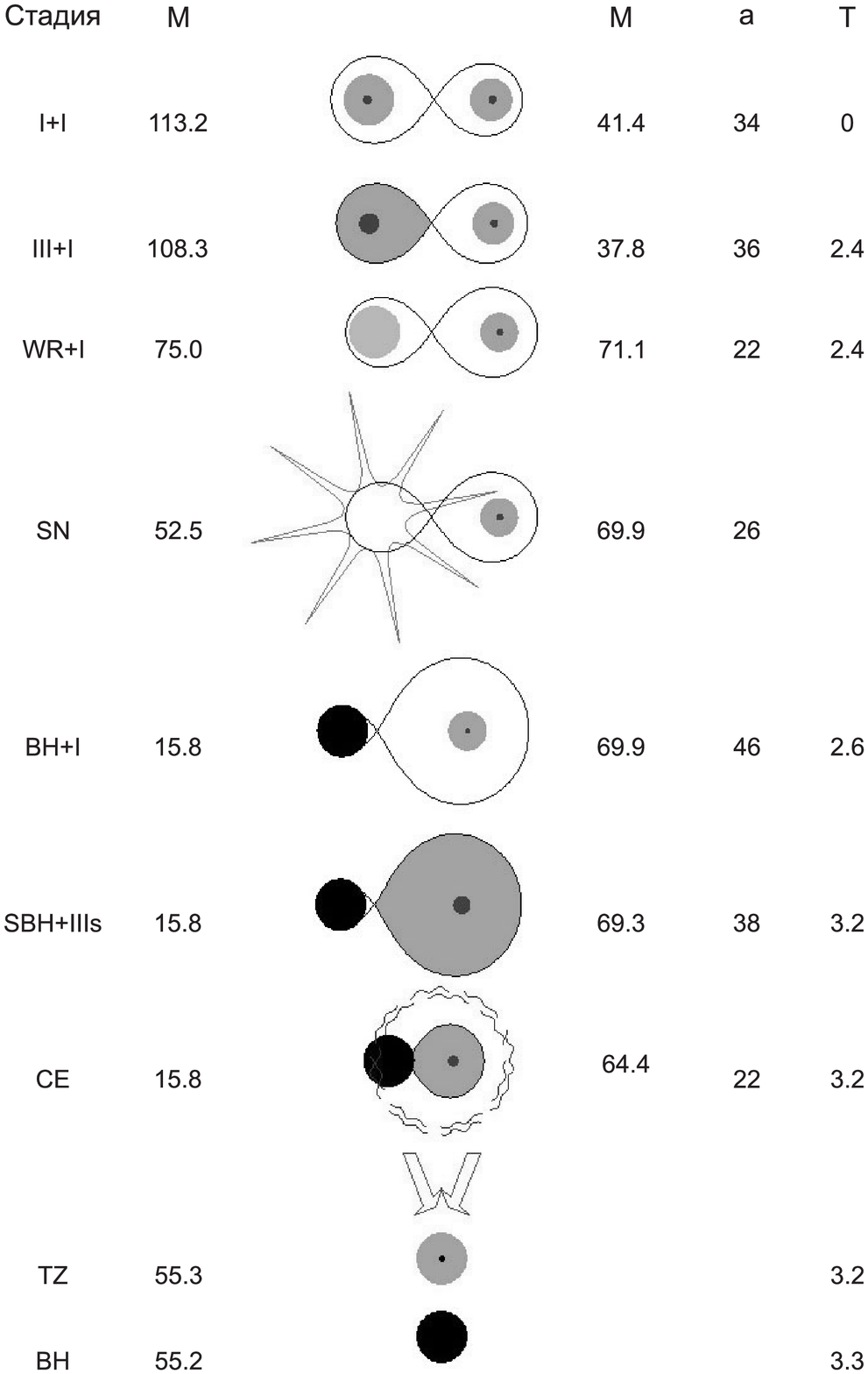} \vspace{0cm}\caption{A characteristic
evolutionary scenario resulting in the formation of an M33
X-7-type system. The notation for the evolutionary stages is (see
\cite{scm2}): for more detail): I main sequence stage, III, IIIs
stage of Roche lobe filling, WR Wolf-Rayet stage, BH black hole,
SBH black hole with a supercritical accretion rate, SN supernova
explosion,CE common envelope stage, TZ Thorne-Zhytkov object. The
mass of the primary and secondary stars $M$ (in $M_{\odot}$), the
semimajor axis $a$ (in $R_{\odot}$), and the age of the system $T$
(in millions of years) are presented for the onsets of the
corresponding stages and, in the case of the supernova explosion,
for the time just before the explosion. The free parameters of the
problem for this scenario are $\alpha=0.3$, $k_{WR}=0.3$, and
$k_{bh}=0.3$.} \label{evol}
\end{figure*}

\begin{figure*}[h!]
\hspace{0cm} \epsfxsize=0.9\textwidth\centering
\epsfbox{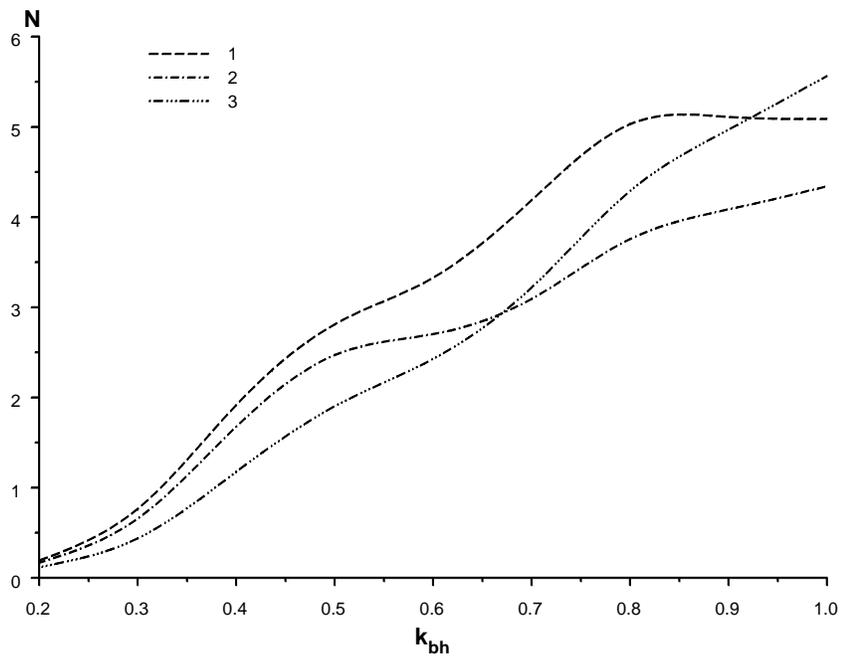} \vspace{0cm}\caption{Same as Fig. 8
for an IC 10 X-1-type system.} \label{numic10}
\end{figure*}

\begin{figure*}[h!]
\hspace{0cm} \epsfxsize=0.5\textwidth\centering
\epsfbox{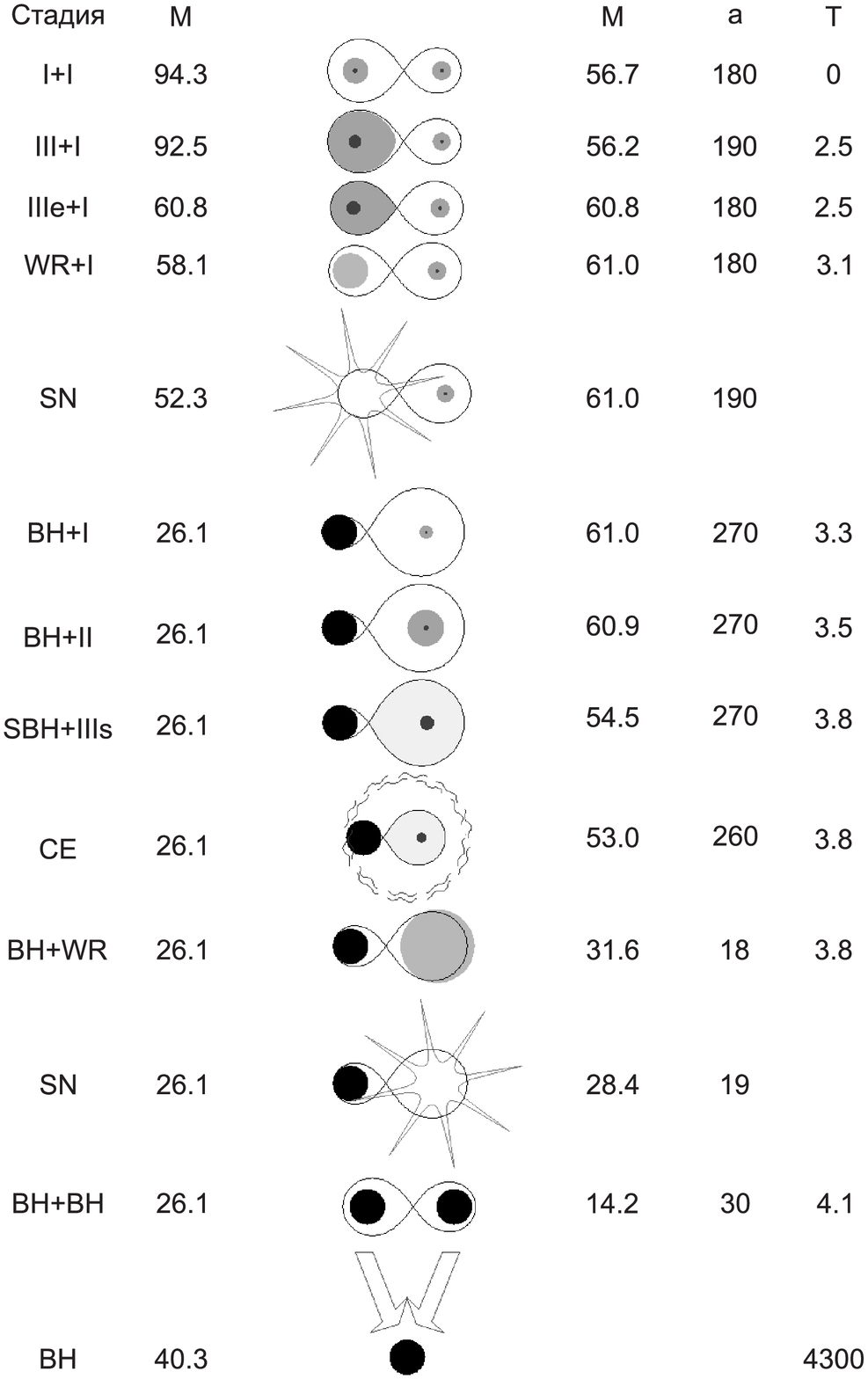} \vspace{0cm}\caption{A characteristic
evolutionary scenario resulting in the formation of an IC 10
X-1-type system. The notation for the evolutionary stages is (see
\cite{scm2}): for more detail): I main sequence star; II star has
ended its evolution on the main sequence but does not fill its
Roche lobe; III, IIIe, and IIIs stage of Roche lobe filling; WR
Wolf-Rayet stage; BH black hole; SBH black hole with a
supercritical accretion rate; SN supernova explosion; CE common
envelope stage. The mass of the primary and secondary stars $M$
(in $M_{\odot}$), the semimajor axis $a$ (in $R_{\odot}$), and the
age of the system $T$ (in millions of years) are presented for the
onsets of the corresponding stages and, in the case of the
supernova explosion, for the time just before the explosion. The
free parameters of the problem in this scenario are $\alpha=0.1$,
$k_{WR}=0.1$, and $k_{bh}=0.5$.} \label{evolic10}
\end{figure*}

\begin{figure*}[h!]
\hspace{0cm} \epsfxsize=0.9\textwidth\centering
\epsfbox{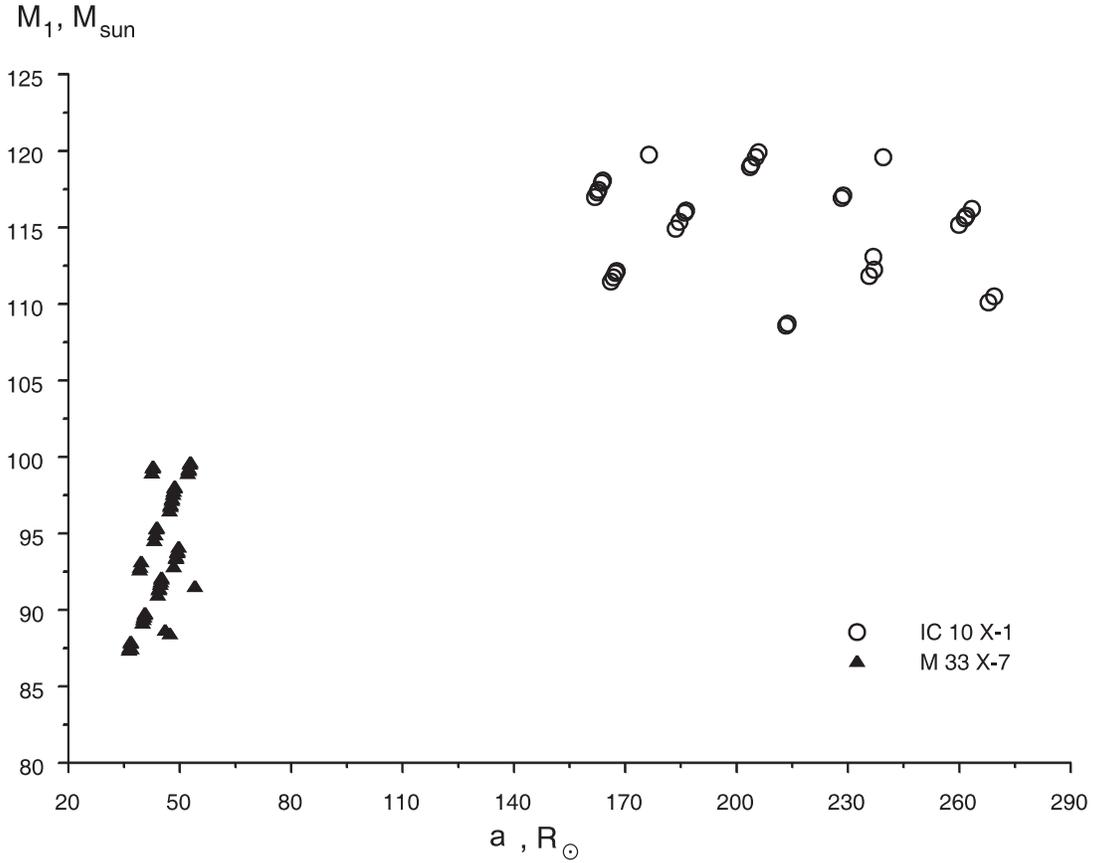} \vspace{0cm}\caption{Initial parameters of
systems that evolve into binaries of the type considered. The
vertical axis plots the initial mass of the primary (which is
initially more massive) and the horizontal axis the initial
semimajor axis of the system. The triangles and circles indicate
binaries that evolve to form M33 X-7-type and IC 10 X-1-type
systems, respectively. The free parameters of the problem in this
case are $\alpha=0.1$, $k_{WR}=0.1$, $k_{bh}=0.3$.} \label{m-a}
\end{figure*}

\begin{figure*}[h!]
\hspace{0cm} \epsfxsize=0.9\textwidth\centering
\epsfbox{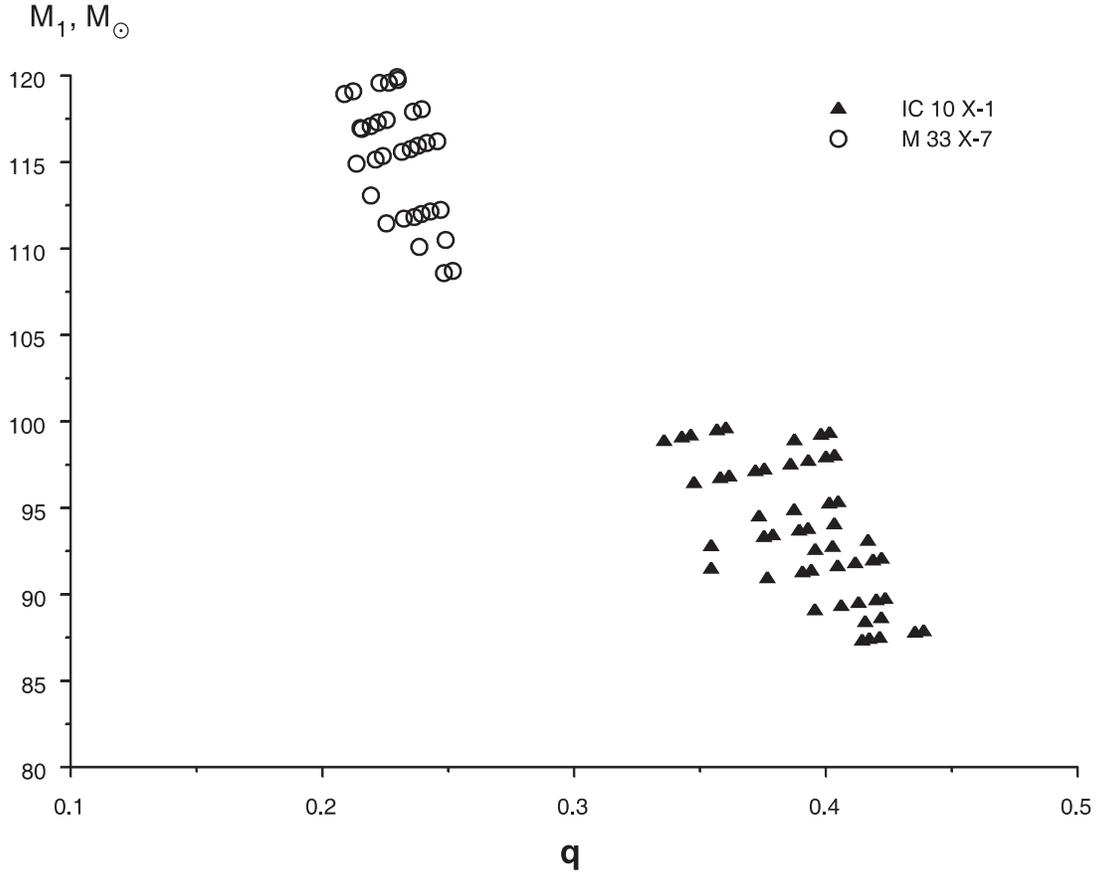} \vspace{0cm}\caption{Initial parameters of
systems that evolve into binaries of the type considered. The
vertical axis plots the initial mass of the primary (which is
initially more massive) and the horizontal axis the component-mass
ratio $q=M_2/M_1<1$ at the beginning of the system's evolution.
The notation is the same as in Fig. 12. The free parameters of the
problem in this case are $\alpha=0.1$, $k_{WR}=0.1$,
$k_{bh}=0.3$.} \label{m-q}
\end{figure*}

\end{document}